\documentclass[%
reprint,groupedaddress,
amsmath,amssymb,aps,prl]{revtex4-2}

\usepackage{graphicx}
\usepackage{dcolumn}
\usepackage{bm}
\usepackage{soul,color}
\usepackage{comment}
\usepackage{lineno}
\usepackage{blindtext}
\usepackage[dvipsnames]{xcolor}
\usepackage{printlen}
\usepackage[citecolor=magenta, colorlinks=true,linkcolor=blue]{hyperref}
\usepackage{xcolor}
\usepackage{pdfcomment}                              
\newcommand{\figref}[2]{Fig.\,\textcolor{blue}{#1#2}}
	\newcommand{\ee}{\end{equation}}

\begin{document}
	\preprint{APS/123-QED}

\title{Non-contact excitation of multi-GHz lithium niobate electromechanical resonators}

\author{Danqing Wang$^{1,\dagger}$, Jiacheng Xie$^{1,\dagger}$, Yu Guo$^{1}$, Mohan Shen$^{1}$, Hong X. Tang$^{1,*}$}

\affiliation{$^1$Department of Electrical Engineering, Yale University, New Haven, Connecticut 06511, USA\\
$^*$hong.tang@yale.edu\\
$^{\dagger}$Authors contribute equally}		
\date{\today}

	\begin{abstract}
	The demand for high-performance electromechanical resonators is ever-growing across diverse applications, ranging from sensing and time-keeping to advanced communication devices. Among the electromechanical materials being explored, thin-film lithium niobate stands out for its strong piezoelectric properties and low acoustic loss. However, in nearly all existing lithium niobate electromechanical devices, the configuration is such that the electrodes are in direct contact with the mechanical resonator. This configuration introduces an undesirable mass-loading effect, giving rise to spurious modes and additional damping. Here, we present an electromechanical platform that mitigates this challenge by leveraging a flip-chip bonding technique to separate the electrodes from the mechanical resonator. By offloading the electrodes from the resonator, our approach yields a substantial increase in the quality factor of these resonators, paving the way for enhanced performance and reliability for their device applications.
	\end{abstract}
	\maketitle
	
\section*{Introduction}
Electromechanical resonators are at the core of many resonator technologies, particularly in communication applications. The advent of fifth-generation (5G) networks has heightened the demand for high-performance electromechanical resonators \cite{le2021piezoelectric}. High frequency resonators in the upper microwave and millimeter-wave bands offer expansive bandwidths for high-data-rate communications. However, the substantial insertion loss at these frequencies presents challenges for scaling electromechanical resonators to operate within this range. Hence, there is a critical need to devise high-frequency, low-loss resonators to address this challenge effectively. 

Among various material platforms of micro-electromechanical systems (MEMS), thin-film lithium niobate (TFLN) has attracted significant attention thanks to its excellent piezoelectric properties \cite{weis1985lithium}, low acoustic loss \cite{shen2020high}, and compatibility with large-wafer thin-film processing. Recently, large electromechanical coupling has been reported with z-cut TFLN lamb wave modes \cite{yang2018toward}. Moreover, sub-THz TFLN electromechanical resonators have also been demonstrated \cite{Xie2023-qy}, showcasing the potential of such a platform for broadband applications. To further enhance TFLN resonators for RF filter applications and quantum phononics research, considerable efforts have been dedicated to improving the quality factor (Q) of devices, including reducing surface roughness of TFLN \cite{Link2021-np}, optimizing the release process \cite{Faizan2021-hs}, and modifying the design and materials of coupling electrodes \cite{Faizan2021-hs,Yang2020-ns,Yang2021-kz}. However, in all these implementations, the metal electrodes are directly in contact with the lithium niobate (LN) membranes, which leads to significant mechanical losses in the electromechanical systems \cite{Tu2020-jw,Kim2018-iu,Wollack2021-wu,Segovia-Fernandez2017-dg}. 

The substantial loss accompanied by the contact electrodes motivates researchers to develop mechanical resonators excited through non-contact electrodes. Ref. \cite{Yen2013-hg} employed capacitive-piezoelectric aluminum nitride (AlN) lamb wave resonators to showcase the Q enhancement of the non-contact configuration at around 1\,GHz. In this paper, we experimentally demonstrate a non-contact TFLN electromechanical platform that can operate up to around 30\,GHz. By employing a flip-chip bonding approach, we achieve separation of the electrodes from the resonator body through an air gap, resulting in effectively reduction of mechanical losses and suppression of spurious modes. We propose that by further reducing the gap between electrodes and the membrane to a nanometer scale, we could still achieve a notable enhancement of the device quality factor meanwhile maintaining a large electromechanical coupling strength.

\begin{figure}[t]
    \centering
    \includegraphics [width=1\columnwidth]
    {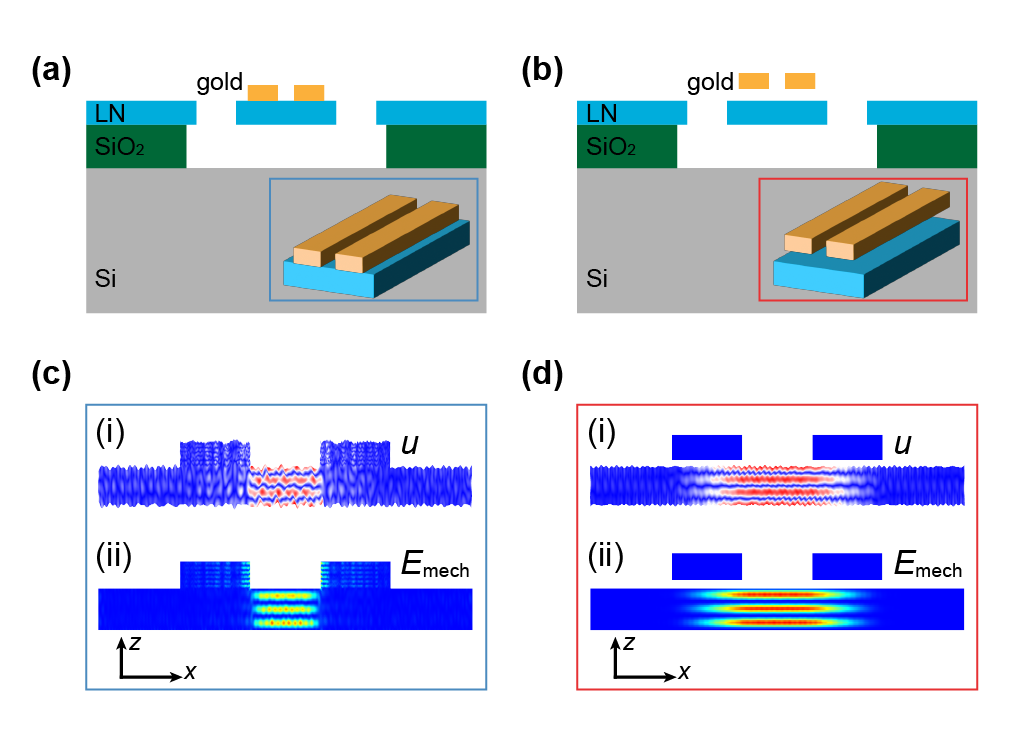}
    \caption{Schematics of direct-contact and non-contact LN electromechanical resonators. (a) Gold electrodes in direct contact with the LN membrane. (b) Electrodes suspended over the LN membrane. Insets of (a)(b) show three-dimensional perspectives of the central region of electromechanical resonators. (c)(d) Simulated on-resonance displacement field (i) and mechanical energy distribution (ii) of the third-order TS mode (TS-3). Color from blue to red maps the displacement amplitude and stored energy from minimum to maximum.}
    \label{fig:1}
\end{figure}

\begin{figure}[t]
    \centering
    \includegraphics [width=1\columnwidth]
    {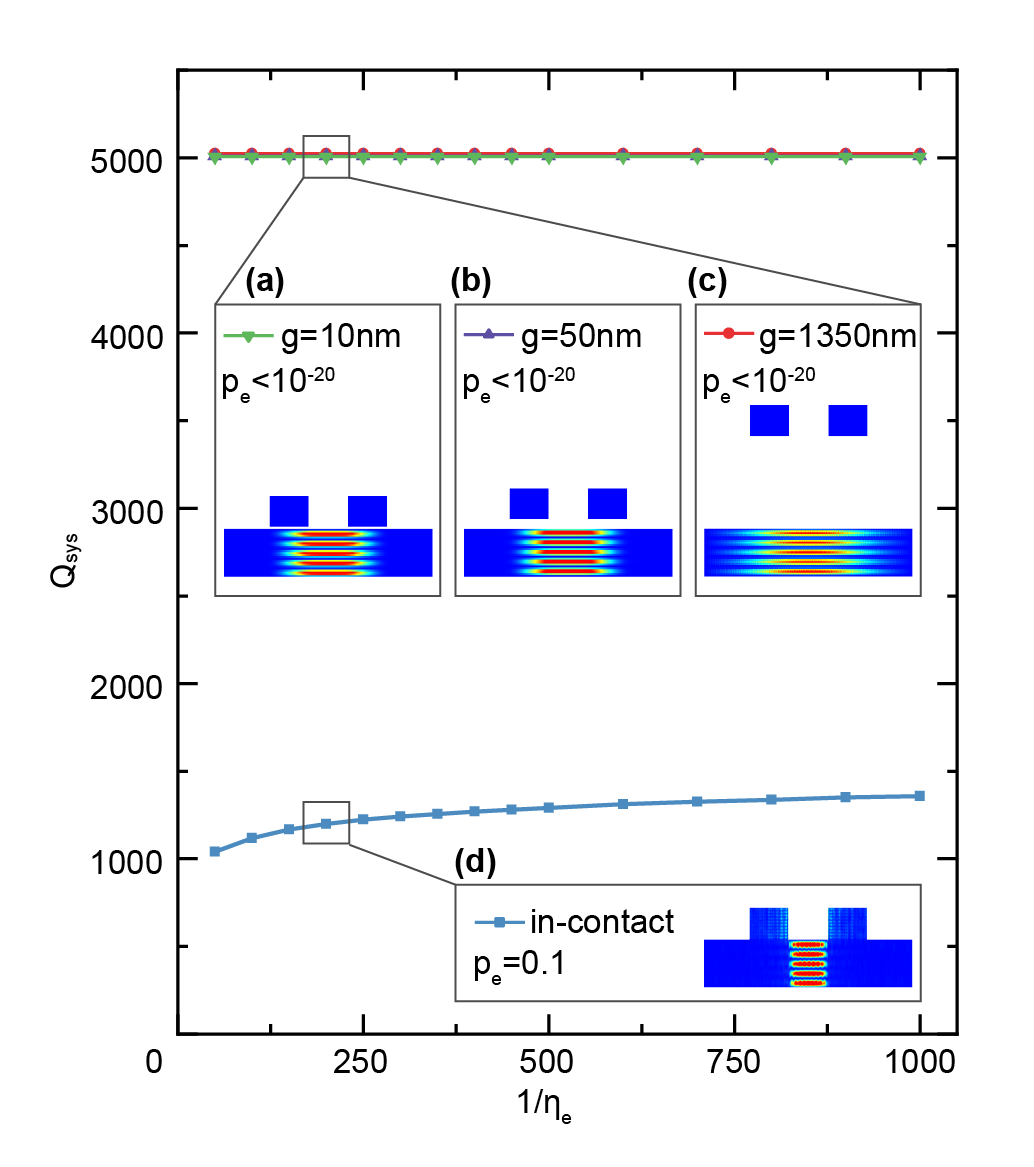}
    \caption{Simulated system Q versus assumed gold loss factor $\eta_\mathrm{e}$. The blue line represents the direct-contact case, while the green, purple, and red lines depict non-contact case with air gaps of 10\,nm, 50\,nm and 1350\,nm, respectively. Simulations are based on the TS-5 mode of 300\,nm TFLN with 200\,nm gold electrodes, and Q is fitted using a multi-resonance mBVD model. (a-d) display each configuration's energy distribution and participation ratio at $\eta_\mathrm{e}$=1/200.}
    \label{fig:2}
\end{figure}

\begin{figure*}[t]
    \centering
    \includegraphics [width=1\linewidth]
    {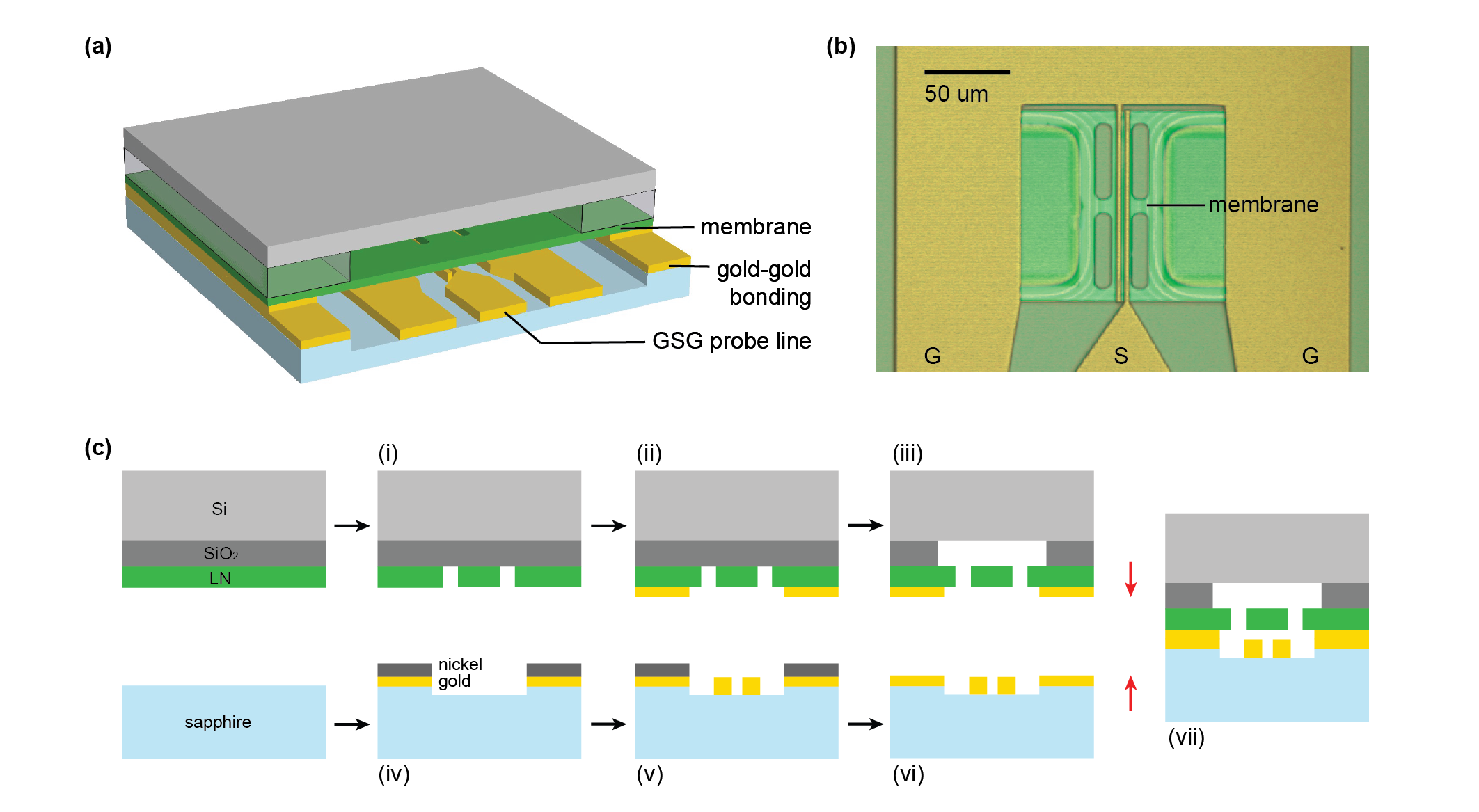}
    \caption{(a) Schematic of a flip-chip bonded non-contact resonator (not to scale). (b) Optical microscope image of a device, viewed from the top electrode side. The width of the membrane is 16\,$\mathrm{\mu}$m and the length is 140\,$\mathrm{\mu}$m. (c) Fabrication flow: (i) LN is etched by Ar ion milling. (ii) gold is deposited at the bonding area. (iii) SiO$_2$ buffer layer is removed in BOE and the chip is dried in CPD. (iv) gold and nickel are deposited at the bonding area of the sapphire chip, and the remaining area is etched around 150\,nm. (v) 200\,nm-thick gold electrodes are deposited. (vi) nickel is removed. (vii) two chips are bonded.}
    \label{fig:3}
\end{figure*}

\begin{figure*}[t]
    \centering
    \includegraphics [width=1\linewidth]
    {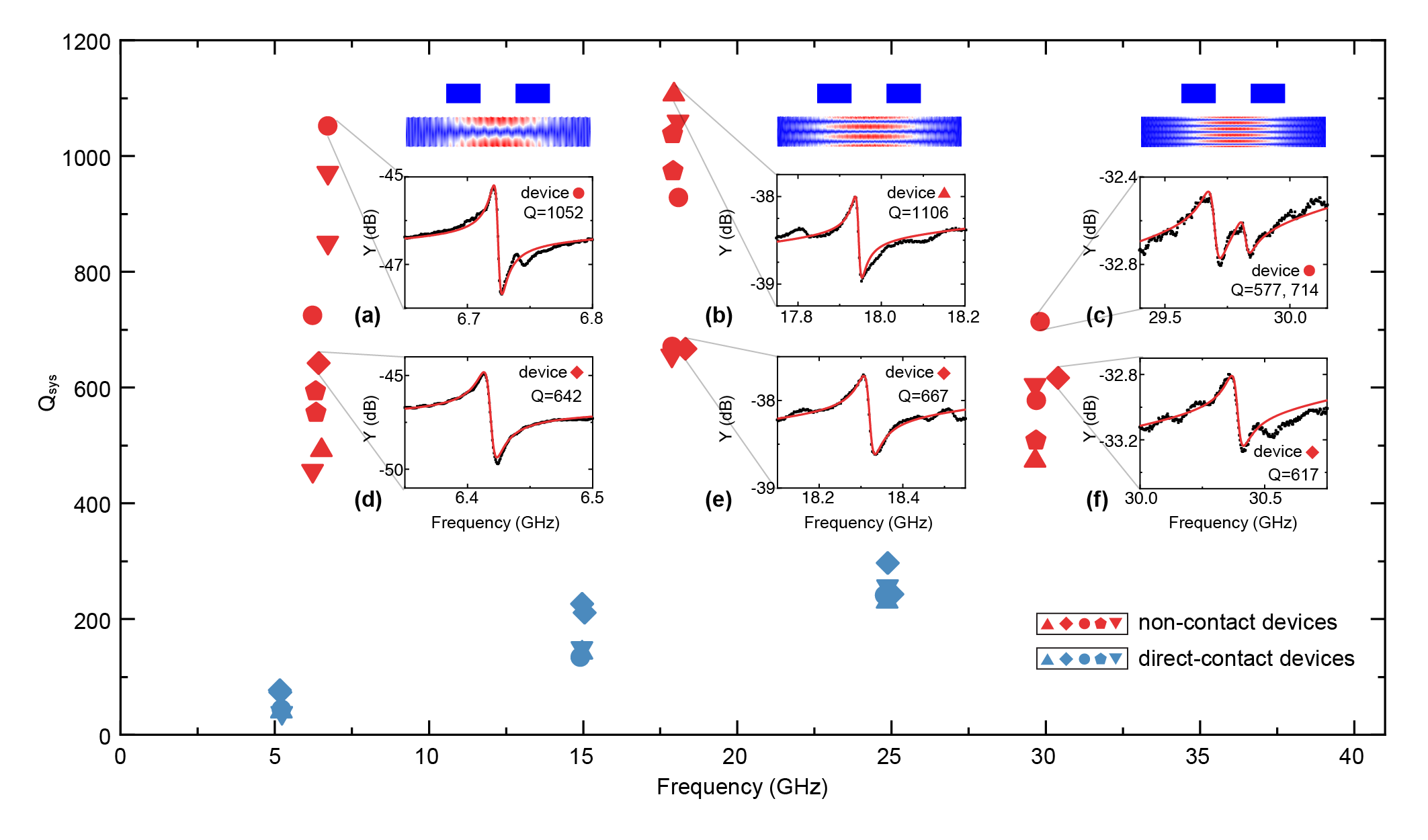}
    \caption{Measured Q versus modal frequency for non-contact devices (red symbols, LN thickness $\sim$300\,nm) and direct-contact devices (blue symbols, LN thickness $\sim$370\,nm). Five devices from each category are marked as up-triangle, diamond, circle, pentagon and down-triangle. Insets (a-f) plot admittance spectra near the first three odd TS modes. The raw data, presented as black dots, are fitted by the mBVD model as red curves. (a-c) show the data from devices with highest Q for each mode. (d-f) represent the spectra from a chosen device (red diamond).}
    \label{fig:4}
\end{figure*}

\section*{Device Design and Fabrication}
As depicted in \figref{\ref{fig:1}}, panel \textcolor{blue}{(a)} illustrates the conventional configuration where the electrodes make direct contact with the LN membrane, while panel \textcolor{blue}{(b)} outlines our proposed scheme where an air gap is introduced between the electrodes and the membrane. In these configurations, a horizontal electric field is utilized to couple to the $z$-cut TFLN thickness-shear (TS) modes \cite{Xie2023-qy} through the large $e_{51}$ piezoelectric coupling element. These thickness modes, bounded by the free top and bottom surfaces of the membrane, have a near-dispersionless characteristic where the mechanical resonant frequency $f$, mode order $m$, film thickness $h$, and acoustic velocity $v$ are related by $mv=2hf$. 

In the direct-contact system, acoustic waves inevitably propagate into the electrodes, as indicated by the displacement field $u$ in \figref{\ref{fig:1}}{(c-i)}. Consequently, the electrodes provide a dissipation path for the mechanical energy $E_{\mathrm{mech}}$ as shown in \figref{\ref{fig:1}}{(c-ii)}. In this scenario, mechanical damping associated with the electrodes contributes to the overall decay of the stored energy in the electromechanical resonator. In contrast, the non-contact platform circumvents this issue by fully confining the acoustic energy in the LN membrane, leveraging the significant mechanical impedance mismatch between the solid and the air. This confinement is visualized through the displacement field and the energy distribution in \figref{\ref{fig:1}}{(d-i,-ii)}. 

Loss mechanism inside an electromechanical resonator can be quantitatively described by an energy participation ratio model:
\begin{equation}
    \frac{1}{Q_{\mathrm{sys}}}=\sum p_i\eta_i\approx p_\mathrm{m}\eta_\mathrm{m}+p_\mathrm{e}\eta_\mathrm{e}
\end{equation}
where $\mathrm{Q}_{\mathrm{sys}}$ is the mechanical quality factor of the whole system, $\eta_i$ is the loss factor \cite{carfagni1998loss} of each loss channel and $p_i=E_i/E_\mathrm{{total}}$ is the energy participation ratio of each channel. Here, we focus on the loss induced by the LN membrane (loss factor $\eta_\mathrm{m}$), including LN's intrinsic damping and the anchor loss, and the loss associated with the gold electrodes (loss factor $\eta_\mathrm{e}$). In the non-contact scenario, the mechanical energy confined in the TFLN membrane is dominant while $p_\mathrm{e}$ is negligible, which contributes to a high quality factor. To validate the advantage of the non-contact platform numerically, we perform finite-element method (FEM)-based simulations (COMSOL) to verify its Q-enhancement. As for the damping parameters used in the simulations, we choose a realistic value of $1/5000$ as $\eta_\mathrm{m}$ for the TFLN, based on the extracted data from a LN acoustic delay line \cite{Lu2021-qp}. Considering that there is no unified value of the gold loss factor, we sweep across a broad range of $\eta_\mathrm{e}$ (1/50 to 1/1000 \cite{Segovia-Fernandez2017-dg,Major2013-dx,Sandberg2005-fz}) to validate the superiority of the non-contact platform in all these scenarios. The corresponding system Q of the fifth-order TS mode (TS-5) at 30\,GHz is shown in \figref{\ref{fig:2}}, for four different configurations: (a) 10\,nm air gap ($g$); (b) 50\,nm air gap; (c) 1350\,nm air gap; (d) electrodes in contact with the membrane. In the direct-contact device, gold electrodes will vibrate with the LN membrane, as shown in \figref{\ref{fig:2}}{(d)}. For all loss factor $\eta_\mathrm{e}$ considered, the energy participation ratio of the electrodes ($p_\mathrm{e}$) ranges from 4\% to 40\%, which is a limiting factor of the mechanical Q of the system. Moreover, spurious modes introduced by electrodes broaden the resonant spectrum, leading to a decrease in the observed system Q. However, in the non-contact platform, as shown in \figref{\ref{fig:2}}{(a-c)}, the air gap prevents acoustic waves from propagating into the electrodes, resulting in a significantly reduced $p_\mathrm{e}$ (less than $10^{-20}$, obtained from simulation), and the system Q is close to the intrinsic value of the LN, regardless of the size of the air gap, demonstrating advantages over the direct-contact platform.

We employ a flip-chip bonding technique to realize a narrow air gap between the TFLN membrane and the electrodes. The complete resonator construction is depicted in \figref{\ref{fig:3}}{(a)}, where a sapphire chip with metal electrodes is bonded to a silicon chip with suspended LN membranes using a gold-gold bonding technique \cite{Higurashi2017-sc}. The fabrication process is outlined in \figref{\ref{fig:3}}{(c)}. For the LN membrane chip, we begin with a chip comprising LN(300\,nm) and SiO$_2$(5\,$\mu$m) on a silicon substrate. The release windows are defined in the LN layer by patterning HSQ resist through electron-beam lithography, followed by Ar ion milling (\figref{\ref{fig:3}}{(c-i)}). A bonding gold layer is then deposited through a lift-off process (\figref{\ref{fig:3}}{(c-ii)}). Subsequently, the membrane chip is released in buffered oxide etchant (BOE) which isotropically removes the SiO$_2$ layer beneath. The released membrane chip is dried by a critical point dryer (CPD) (\figref{\ref{fig:3}}{(c-iii)}). For the electrode chip, we start with a bare sapphire substrate. We opt for sapphire due to its excellent microwave properties and transparency that aids in flip-chip alignment. Typically, the surface roughness of the bonding gold is required to be sub-nm and therefore, its thickness is controlled to be within the range of several tens of nm to achieve the best bonding result. In this step, we deposit 40\,nm gold as the bonding layer. Given that this layer is thinner than the gold electrodes with 200\,nm thickness in our design, we compensate for the thickness difference by creating a recessed area in sapphire to accommodate the electrodes (\figref{\ref{fig:3}}{(c-iv)}). This step is achieved by using Cl$_2$ and BCl$_3$ plasma etching, with nickel as the hard mask. Then, 200\,nm-thick gold electrodes are deposited using a lift-off technique with Copolymer/PMMA resists (\figref{\ref{fig:3}}{(c-v)}). The nickel mask is finally removed in a piranha solution before bonding (\figref{\ref{fig:3}}{(c-vi)}). In the final step, we use a commercial bonding tool to bond the LN membrane chip and the electrode chip, resulting in a well-defined air gap between the membrane and electrodes (\figref{\ref{fig:3}}{(c-vii)}). The optical microscope image of a completed flip-chip bonded device is shown in \figref{\ref{fig:3}}{(b)}. 

\section*{Results and discussion}
A ground-signal-ground (GSG) probe is employed to engage with the electrodes on the sapphire chip for measuring the non-contact resonator. The electrode pads on the sapphire chip are designed to protrude beyond the edge of the TFLN, facilitating the probe access, as shown in \figref{\ref{fig:3}}{(a)}. Via LN's piezoelectricity, the mechanical mode is coupled to the microwave field, and can be read out through a reflection measurement using a vector network analyzer.

The admittance spectra near the first three odd TS modes are plotted in the insets \textcolor{blue}{(a-f)} of \figref{\ref{fig:4}}. A modified Butterworth-Van Dyke (mBVD) model is used to fit the experimental data \cite{larson2000modified}, represented by solid lines in the insets of \figref{\ref{fig:4}}. In the mBVD model, the mechanical resonator is depicted as the combination of three lumped elements in series ($Z_\mathrm{mech}={1}/({\mathrm{i}\omega C_\mathrm{mech}})+\mathrm{i}\omega L_\mathrm{mech} + R_\mathrm{mech}$). In addition to exciting the resonator, two electrodes are coupled with each other, equivalent to a mutual capacitor ($C_\mathrm{M}$). The mechanical Q of devices is extracted as $\sqrt{L_\mathrm{mech}/C_\mathrm{mech}}/R_\mathrm{mech}$. The Q values of multiple mechanical resonators excited by both non-contact and direct-contact electrodes are plotted in the main graph of \figref{\ref{fig:4}}, which demonstrates significant Q enhancement of the non-contact platform. Notably, the highest experimental Q values are 1052 for TS-1 mode, 1106 for TS-3 mode and 714 for TS-5 mode, as depicted in \figref{\ref{fig:4}}{(a-c)}. To illustrate systematic Q-improvement of the non-contact devices, we additionally plot the admittance spectra of a chosen device with average performance, as shown in \figref{\ref{fig:4}}{(d-f)}. Remarkably, the Q values of the three modes, all exceeding 600, are still substantially higher than those observed in the direct-contact system. Therefore, despite the variations of the measured Qs in each category, it is clear that the non-contact platform systematically achieves a higher Q performance than the direct-contact case. Moreover, as indicated by the simulations in \figref{\ref{fig:2}}, this non-contact technique for enhancing the Q is universally applicable to piezoelectric materials with various loss properties.

To estimate the achieved air gap distance, we image the membrane chip using a 3D optical profilometer and observe a buckling of approximately 1.3\,$\mathrm{\mu}$m in the suspended 300\,nm-thick LN film toward the silicon substrate. The extent of this buckling is consistent with the optical interference patterns observed on the suspended TFLN membrane, as shown in \figref{\ref{fig:3}}{(b)}. The buckling of the membrane could derive from the intrinsic compressive stress of the LN films and is under further investigation. Since the resulting gap distance exceeds the designed value, the electromechanical coupling coefficients ($K^2$, defined as $C_\mathrm{mech}/C_\mathrm{M}$) are lower than expected, measuring 0.02$\%$ and 0.01$\%$ for the TS-3 and TS-5 modes of the device represented by the red diamond symbol in \figref{\ref{fig:4}}, respectively. According to FEM simulations, if the air gap reaches 50\,nm as designed, the $K^2$ of the non-contact case are 2.5$\%$ and 0.9$\%$ for the TS-3 and TS-5 modes, both of which are around 70\% of those in the direct-contact case. Consequently, addressing the buckling in LN thin films will be a primary focus in the upcoming stages of development. The non-contact configuration with reduced buckling will deliver high performance in both achievable quality factors and $K^2$ characteristics.

\section*{Conclusions}
We develop a flip-chip bonding technique to excite LN resonators by non-contact electrodes, resulting in notable improvement of the mechanical Q compared to the direct-contact case in the multi-GHz frequency regime (up to 30\,GHz). This configuration not only effectively reduces the loss associated with the electrodes, but also holds the potential to maintain a high electromechanical coefficient. The device structure demonstrated here can be extended to other piezoelectric materials, such as AlN. With further development, the fabrication process demonstrated in this work will permit the fabrication of even thinner nanomembranes below 100\,nm. These collective advantages make non-contact electromechanical resonators well-suited for high-frequency applications.

	\section*{References}
	\bibliography{references}
	
\vspace{8mm}\noindent{\bf Acknowledgements}\\
This project is supported in part by the Air Force Office of Sponsored Research (AFOSR MURI FA9550-23-1-0338) and the Defense Advanced Research Projects Agency (DARPA OPTIM HR00112320023). The part of the research that involves lithium niobate thin film preparation is supported by the US Department of Energy Co-design Center for Quantum Advantage (C2QA) under Contract No. DE-SC0012704. The authors would like to thank Yong Sun, Lauren Mccabe, Kelly Woods, and Michael Rooks for their assistance provided in the device fabrication. The fabrication of the devices was done at the Yale School of Engineering \& Applied Science (SEAS) Cleanroom and the Yale Institute for Nanoscience and Quantum Engineering (YINQE).	
	
\vspace{1mm}\noindent{\bf Author contributions}\\
H.X.T., J.X., D.W. conceived the experiment. J.X. and D.W. designed the devices. D.W. did the fabrications with contribution from Y.G., J.X. and M.S. J.X. and D.W. performed the measurement. D.W. and J.X. did simulations and processed the data with contribution from M.S..  D.W., J.X. and H.X.T. wrote the manuscript with input from all the authors. H.X.T. supervised the project.

\vspace{1mm}\noindent{\bf Competing interests}\\
The authors declare no competing interests.	
\end{document}